# Proteins with alternative folds reveal blind spots in AlphaFold-based protein structure prediction

Devlina Chakravarty[1], Myeongsang Lee[1], and Lauren L. Porter[1,2*]

[1]National Center for Biotechnology Information, National Library of Medicine, National Institutes of Health, Bethesda, MD 20894
[2]Biochemistry and Biophysics Center, National Heart, Lung, and Blood Institute, National Institutes of Health, Bethesda, MD, 20892
*Correspondence: porterll@nih.gov

**Abstract**

In recent years, advances in artificial intelligence (AI) have transformed structural biology, particularly protein structure prediction. Though AI-based methods, such as AlphaFold (AF), often predict single conformations of proteins with high accuracy and confidence, predictions of alternative folds are often inaccurate, low-confidence, or simply not predicted at all. Here, we review three blind spots that alternative conformations reveal about AF-based protein structure prediction. First, proteins that assume conformations distinct from their training-set homologs can be mispredicted. Second, AF overrelies on its training set to predict alternative conformations. Third, degeneracies in pairwise representations can lead to high-confidence predictions inconsistent with experiment. These weaknesses suggest approaches to predict alternative folds more reliably.

**Introduction**

Deep learning models have revolutionized protein structure prediction. AlphaFold2 (AF2), developed by DeepMind, is a state-of-the-art neural network that accurately predicts many protein structures from their amino acid sequences, including some outside of its training set [1]. AF2's remarkable ability to extract detailed structural information from multiple sequence alignments (MSAs) and convert it to a structure has enabled it to surpass the results of its predecessors [2,3]. Furthermore, the newly released AlphaFold3 (AF3) employs a refined neural network architecture with a multiscale diffusion process to predict protein-protein interactions and protein-ligand complexes, including nucleic acids and ions with impressive accuracy [4]. These and numerous other approaches have established the utility of deep learning approaches in protein structure prediction and design [5-8].

Despite the many successes of these new deep-learning-based methods for protein structure prediction and design, challenges remain. In this review, we focus on protein structures that AlphaFold (AF) struggles to predict accurately: those that assume conformations distinct from their training-set homologs and alternative conformations of fold-switching proteins, which remodel their secondary and/or tertiary structures and change their functions in response to cellular stimuli [9]. We close by suggesting ways to advance predictions of alternative protein conformations.

**Homologs with different structures**

For decades, it has been assumed that homologous protein sequences fold into similar structures [10]. However, as sequence databases have become more populated and more protein structures have been determined, it has been realized that sometimes homologous proteins can evolve distinct folds [11,12]. This complicates protein structure prediction. Similar amino acid sequences typically have similar MSAs with similar evolutionary information used to infer three-dimensional structure. In cases when sequence and structure space blend together [13], it is difficult–though not impossible [11]–to discriminate between distinct conformations using MSAs.

Thus, when homologous sequences assume different structures, it is not surprising that AF2 sometimes incorrectly predicts that a given sequence assumes the structure of its alternatively folded homolog (**Figure 1**). For instance, BCCIP, a human DNA-repair protein linked to cancer development, has two isoforms, α and β, differing by one C-terminal exon substitution [14]. Though the sequences of these two isoforms are ~80% identical, their structures differ by > 12Å. While BCCIPβ is conserved widely among eukaryotes, BCCIPα has been found only among primates and a few other species. AF2/3 predict the structure of BCCIPβ accurately and with high confidence. Unfortunately, they mispredict that BCCIPα assumes the same structure as BCCIPβ. BCCIPα assumes a unique fold: its most similar structure in the Protein Data Bank (PDB) differed by 10Å [14], and its structure could not be in AF2 or AF3's training sets since it was released in 2023. Further, including BCCIPα's binding partner, FAM46A, has no effect on its predicted structure [15]. Thus, it seems that AF2 and AF3 associate the sequence patterns detected for BCCIPα with BCCIPβ-like structures in their training sets. Similarly, human pro-interleukin-18, assumes a structure different from its mature form lacking its 36-residue N-terminal pro-domain [16]. Both AF2 or AF3 confidently predict that pro-interleukin-18 assumes the same structure as the mature form, which was likely in their training sets. Further, the structure of DZZB, an ancestor of transcription and translation regulators, was recently solved and found to assume a unique dimeric topology [12]. Instead of predicting this topology, both AF2 and AF3 predicted that DZZB folds into the structure of its close homolog RIFT, whose fold was likely in the AF2-multimer and AF3's training sets. Finally, both AF2 and AF3 incorrectly predict the oligomeric assembly of the human lens protein MP20: side-by-side orientations of the monomeric units [17] resembling a homolog likely in the training set (PDB ID: 5WEO, released 2017), while experiments show a stacked orientation that has not been observed in the PDB [17].

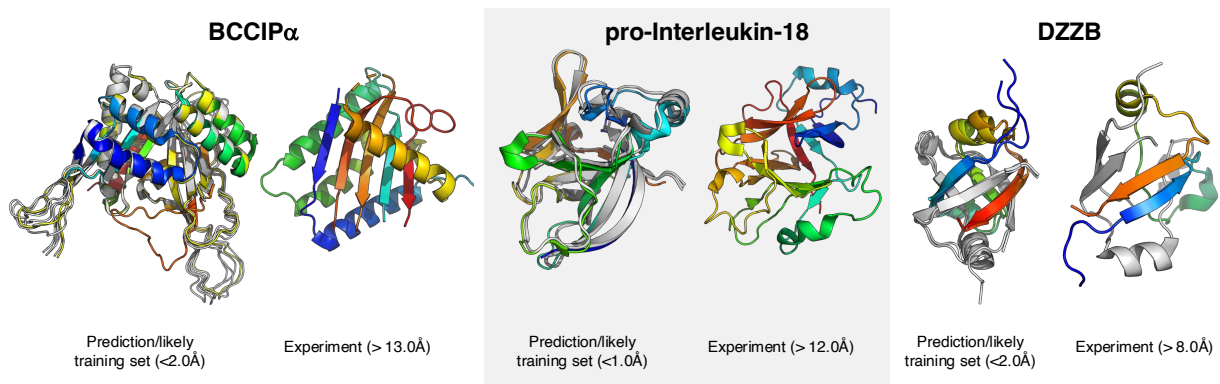

**Figure 1. AlphaFold can struggle to predict structures of proteins with differently folded homologs in its training set.** The human cancer protein BCCIP has two isoforms, α and β, that have completely different folds (left, PDB IDs: 7kys, 8exe chain B, respectively), human pro-interleukin-18, assumes a structure different from its mature form (middle, PDB IDs: 3wo2, 8urv, respectively), and DZZB, an ancestor of transcription and translation regulators, assumes a unique dimeric topology than its training-set homologs (right, PDB IDs: 7dxr, 8jvp, respectively). Training set structures (left, gray) are superimposed on AF3 predictions (left RMSDs) and compared with experiment (colored rainbow from N- to C-termini). AF2 generates similar results as AF3. MP20 not shown because its coordinates were not available. Structures in AF2's/AF3's training sets were deposited before August 29, 2018/October 1, 2021.

## Alternative conformations of proteins

Most experimentally identified proteins exhibit a single dominant folded structure; however, proteins can adopt multiple conformational states within the cellular environment [18]. Cellular stimuli such as thermal fluctuations, ionic strength, small molecules, and binding proteins create distinct energy landscapes that influence protein structure and function. For example, the functionality of allosteric and kinase proteins is contingent upon the active or inactive status of specific loop regions [19-21]. Fold-switching proteins assume two distinct conformations characterized by remodeling their secondary structures [9]. Experimental techniques, including X-ray diffraction and nuclear magnetic resonance (NMR), can capture diverse protein conformations [22,23], although technical limitations–such as sample instability–can impede their characterization. Computational techniques, such as molecular dynamics simulations, can explore various energy landscapes of proteins [24,25], but they require substantial computational resources and time since fold switching occurs on a slow timescale on the order of seconds [26].

AlphaFold2 (AF2) is well-known for predicting single three-dimensional structures of proteins by energy-minimizing evolutionary restraints from MSAs [27]. While evolutionary information derived from deep MSAs often informs predictions of dominant protein structures, these MSAs often leave AF struggling to detect signals for alternative conformations. To overcome this barrier, MSA modifications have been proposed. For instance, Stein and Mchaourab developed a method called SPEACH-AF that predicts alternative conformations by introducing alanine mutations at specific columns within the MSA [28], masking dominant coevolutionary signals. This approach successfully predicted the structures of adenylate kinase (AK), ribose-binding protein (RBP), and several membrane transport proteins. Another notable approach, AFsample2, combined random MSA masking with neural network layer dropout, generating accurate predictions of alternative conformations for transport proteins [29]. Additionally, MSA subsampling has emerged as a viable method for predicting alternative protein conformations [30]. Monteiro da Silva et al. successfully predicted different conformations of Abl kinase and granulocyte-macrophage colony-stimulating factor through MSA subsampling achieving >80% accuracy by NMR [19]. Further, the AF2-RAVE approach, developed by the Tiwary group, showed that predictions derived from reduced MSAs were comparable to those obtained from unbiased molecular dynamics simulations [20].

**Predicting fold-switched conformations**

Fold-switching proteins encode two or more folded states in a single amino acid sequence [31] and either interconvert between two conformations or switch in response to external stimuli [9]. The conformational changes that fold switchers undergo vary from local remodeling to a dramatic transition from α-helix to β-sheet; some fold-switching events are irreversible–such as membrane-inserted pore formation–while others swap domains (**Figure 2**). Fold-switching proteins pose a unique challenge for protein structure prediction models that are based on the one-sequence-one-structure paradigm. Thus, it is not surprising that running AF2 with standard settings tends to predict the dominant conformations of fold switchers only [32].

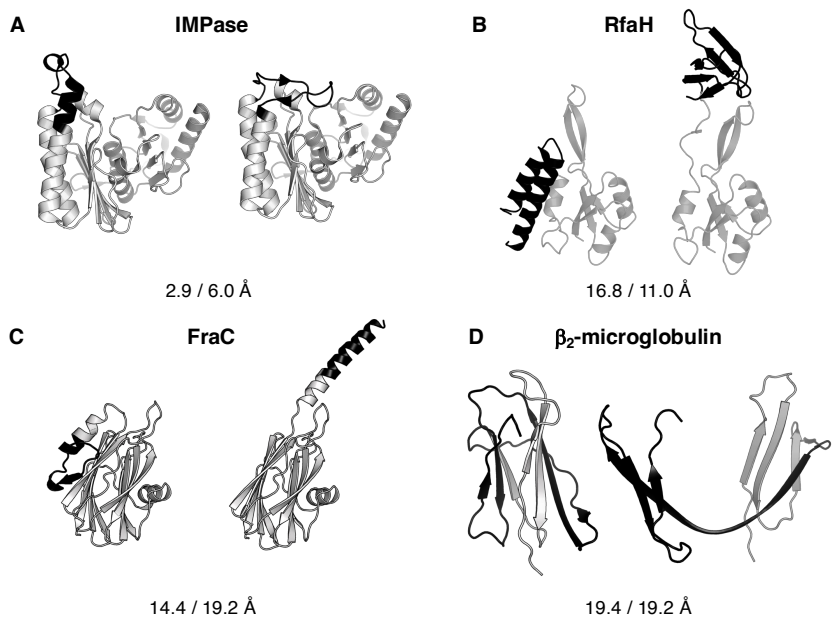

**Figure 2. Examples of conformational changes in fold switchers -** **(A)** An active-site loop of the inositol 1-phosphate phosphatase (IMPase) complex populates both a β-hairpin and **α-**helical structures (PDB ID: 2p3v, chains A and D). **(B)** The C-terminal domain of RfaH changes completely from α-hairpin (PDB ID: 2oug) to a β-roll (PDB ID: 6c6s, chain D) upon binding RNA polymerase and DNA. **(C)** The first 29 residues of the pore-forming fragaceatoxin C (FraC) haemolytic protein switch folds in the presence of lipids and transition from a monomer (left, PDB ID: 3zwg) to an octomer (right, though one subunit is shown for clarity, PDB ID: 4tsy), **(D)** $β_2$-microglobulin, the light chain of the major histocompatibility complex, forms a domain swapped dimer (right, PDB ID: 3low, single subunit shown for clarity) on its pathway from monomer (left, PDB ID: 3m1b, chain F) to pathogenic amyloid fibril. Fold-switching/single folding regions are black/gray. Whole / fold-switching RMSDs for each pair are reported to showcase the spread of conformational changes observed in fold switchers.

In the last few years, several different pipelines have been proposed to diversify the ensembles predicted by AF2 [33]. These methods often rely on reducing the number of sequences in the MSA. The idea behind using a shallower MSA is to weaken co-evolutionary signals from the dominant fold, enabling alternative conformations to be predicted. However, shallower MSAs (especially those with <30 sequences) may produce less accurate models [1].

Though it has been proposed that AF2 predicts alternative conformations of fold switchers by combining coevolutionary inference with a learned energy function [27,33,34], we and others hypothesize that it uses sophisticated pattern recognition to relate input sequences to structures learned during training [15,35,36]. Supporting our hypothesis, after systematically benchmarking all versions of AlphaFold (AF2, AF2_multimer, and AF3) and two AF-based methods that use enhanced sampling to predict alternative conformations of fold-switching proteins (~300,000 proteins total), we found that AF was a weak predictor of fold switching, sampling both conformations in only 35% of the fold switchers that were likely in its training set [15]. Notably, a recent computational method called Alternative Contact Enhancement (ACE), developed in our lab, identified coevolutionary information unique to both folds of 56 fold-switching proteins, confirming that multiple sequence alignments (MSAs) often contain structural information specific to both conformations of fold switchers [31].

These results indicated that the AF2-based enhanced sampling methods evaluated did not seem to leverage the dual-fold coevolutionary information present in the MSAs of fold-switching proteins [15,37]. We also observed that 30-49% of predictions from these enhanced sampling methods did not match either experimentally determined structure. Furthermore, AF2's confidence metrics tended to favor less diverse conformations, assigning lower confidence to accurately predicted diverse conformers and higher confidence to unobserved predictions. Finally, we found strong evidence that AF2 has memorized structural information during its training. For instance, while the coevolutionary information from RfaH's full MSA show strong signals unique to its β-sheet conformation, AF2 predicts its helical form after 3 recycles. Importantly, a version of AF2 recently trained without alternative conformations [35], including the α-helical conformation of RfaH, predicted only the β-sheet conformation from its full MSA [38] (**Figure 3A**).

These findings have significant implications: without coevolutionary information or a robust learning of protein energetics, AlphaFold is often limited to sampling the alternative conformations it has encountered during training, which may help to explain why it struggles to predict some new folds (**Figure 1**). However, random sequence sampling at very shallow depths might enable more reliable predictions for certain fold-switching proteins, suggesting that the memorization and sequence associations in AF2 could facilitate robust sampling of alternative predictions efficiently [30,37].

A recent study by Bryant and Noé [35] explored the challenge of predicting alternative protein conformations using neural network-based methods. To avoid memorization of alternative conformations, they developed a structure prediction network called CFold, built on the AF2 architecture but trained exclusively on a conformational split of the PDB omitting all alternative conformers from the training dataset. To generate alternative conformational samples, they then employed dropout and MSA clustering. Their findings indicated no correlation between the internal embedding representations of the network and the structural outputs, suggesting that the sampling of alternative conformations is stochastic, with coevolutionary information potentially constraining the structure to specific outcomes rather than specifying unique structures. They also found that CFold struggles to predict large conformational changes.

When tested on fold-switching proteins, CFold often failed to predict the alternative fold if that conformer was not included in the training set [38]. By contrast, MD-based methods have successfully generated alternative conformations of fold switchers from sparse coevolutionary restraints [39]. This underscores the dependence that deep learning models sometimes have on structures in their training sets. Interestingly, CFold also predicted incorrect alternative conformations with high confidence. These mispredictions seemed to result from degenerate structures with the same pairwise representation. A similar degeneracy was observed with AF3, which misassigned coevolutionary restraints of XCL1 and predicted its dimeric form incorrectly (**Figure 3B**), whereas AF2 accurately predicted it by relying on memorized structures [15]. Thus, it may be that a larger training set enables deep learning models like AF2 to preferentially select experimentally observed conformations over degenerate solutions.

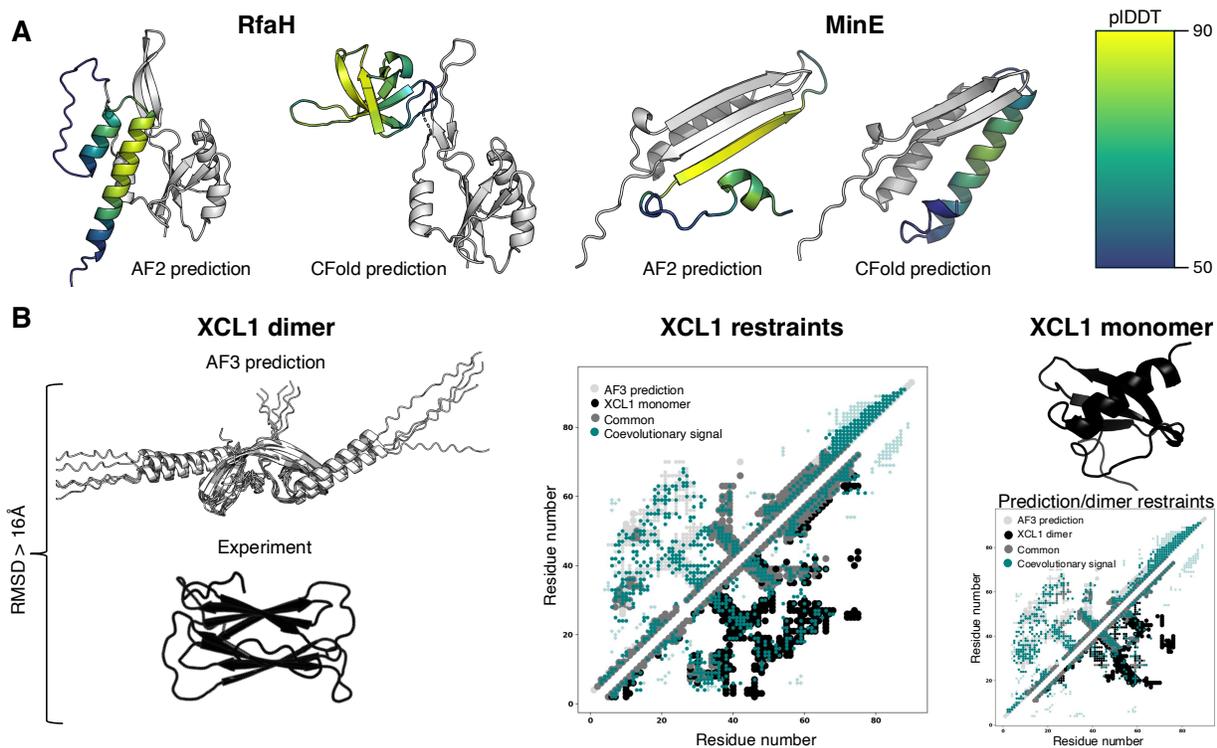

**Figure 3.** Limitations of AlphaFold-based predictions of fold-switching proteins. and misassigns coevolutionary signals. (A). AF2 over-relies on training-set structures. Though RfaH's full MSA has strong coevolutionary signals for its β-sheet form, AlphaFold2 predicts its helical conformation with high accuracy and confidence; the structure of helical RfaH was likely in its training set. CFold, whose training set lacks this helical structure, predicts the β-sheet form with high confidence from the same MSA and sampling depth. Further, AF2 predicts both forms of MinE (likely in its training set) with high accuracy and confidence from its full MSA; the highest confidence prediction–closely resembling its apo dimeric form–is shown above, while CFold predicts only one form from the same MSA with an improperly oriented helix; its training set contains no MinE homologs. (B). AF3 predicts an experimentally inconsistent dimeric form of XCL1 (left). Its contact map (middle) is nearly identical to that of the XCL1 monomer (upper right), whose contact map differs from that of the dimer (bottom right). Coevolutionary signals from MSAs calculated with ACE [31].

## Conclusions

While AI-driven advances have revolutionized predictions of single protein conformations, they do not predict alternative conformations of fold-switching proteins reliably [40]. AF2's success at predicting single folds has been attributed to structural completeness of the PDB [41]. Nevertheless, the protein universe is vast. Some properties of protein structure may not be fully represented in the PDB, as evidenced by newly discovered folds (e. g., **Figure 1**). Previous work has suggested that the PDB is biased toward proteins that are relatively easy to purify and characterize structurally [42,43]. These biases will limit what AF2 can predict since its predictions of fold switchers and large conformational changes are currently driven by its training set [35,38]. Metrics that distinguish between trustworthy and questionable AI-based predictions could guide experimentalists towards discoveries that current AI-based models cannot reveal on their own. This will require modified confidence measures since inaccurate predictions can be scored with high confidence [15,35].

The recent Nobel Prize awarded for AlphaFold recognizes the enormous impact of AI-based methods on protein structure prediction, but there is still work to do. The common pitfalls we have encountered when predicting alternative conformations include: (a) incorrect associations between training-set structures and distinctly folded homologs, (b) overreliance on training-set structures limits which alternative conformations are predicted, and (c) misassigning coevolutionary signals, leading to high-confidence predictions inconsistent with experiment. Addressing these issues will require methods that are more sensitive to mutational effects on protein structure [44]. Further, alternative training approaches that map the same sequence to multiple structures without overfitting are needed [31]. Finally, physically-based priors that eliminate unphysical predictions arising from degenerate contact maps will need to be developed and integrated [45]. These advances will blaze new trails in this frontier of protein structure prediction.

**ACKNOWLEDGEMENTS**

This work utilized resources from the NIH HPS Biowulf cluster (http://hpc.nih.gov) and it was supported by the Intramural Research Program of the National Library of Medicine, National Institutes of Health (LM202011, L.L.P.).

**\*\*A single exon substitution completely changes the structure and alters the function of the human protein BCCIP. AF-based methods struggle to predict one of these conformations.**

15. Chakravarty D, Schafer JW, Chen EA, Thole JF, Ronish LA, Lee M, Porter LL, AlphaFold predictions of fold-switched conformations are driven by structure memorization, Nat Commun 15 (2024) 7296. https://doi.org/10.1038/s41467-024-51801-z.
    **\*\*This study systematically assesses AlphaFold's performance in predicting protein fold switching and finds it to be weak. Several lines of evidence suggest these weaknesses arise from memorization of training-set structures.**
16. **Bonin JP, Aramini JM, Dong Y, Wu H, Kay LE, AlphaFold2 as a replacement for solution NMR structure determination of small proteins: Not so fast!, Journal of Magnetic Resonance (2024) 107725. https://doi.org/10.1016/j.jmr.2024.107725.**
    *A recently solved NMR structure of a small protein does not match its high-confidence AlphaFold predictions. Instead, the predictions closely resemble a training-set structure.
17. Nicolas WJ, Shiriaeva A, Martynowycz MW, Grey AC, Ruma Y, Donaldson PJ, Gonen T, Structure of the lens MP20 mediated adhesive junction, bioRxiv (2024) 2024.2005.2013.594022. https://doi.org/10.1101/2024.05.13.594022.
18. Wei G, Xi W, Nussinov R, Ma B, Protein Ensembles: How Does Nature Harness Thermodynamic Fluctuations for Life? The Diverse Functional Roles of Conformational Ensembles in the Cell, Chem Rev 116 (2016) 6516-6551. https://doi.org/10.1021/acs.chemrev.5b00562.
19. **Monteiro da Silva G, Cui JY, Dalgarno DC, Lisi GP, Rubenstein BM, High-throughput prediction of protein conformational distributions with subsampled AlphaFold2, Nat Commun 15 (2024) 2464. https://doi.org/10.1038/s41467-024-46715-9.**
    *MSA-subsampling methods enabled AF2 to predict numerous experimentally consistent alternative conformations of two proteins.
20. **Vani BP, Aranganathan A, Tiwary P, Exploring Kinase Asp-Phe-Gly (DFG) Loop Conformational Stability with AlphaFold2-RAVE, J Chem Inf Model 64 (2024) 2789-2797. https://doi.org/10.1021/acs.jcim.3c01436.**
    *Physics-based simulations are used to explore conformational equilibria inaccessible to AF2-based methods alone.
21. Faezov B, Dunbrack Jr RL, AlphaFold2 models of the active form of all 437 catalytically-competent typical human kinase domains, bioRxiv (2023) 2023.2007.2021.550125. https://doi.org/10.1101/2023.07.21.550125.
22. Cai M, Agarwal N, Garrett DS, Baber J, Clore GM, A Transient, Excited Species of the Autoinhibited alpha-State of the Bacterial Transcription Factor RfaH May Represent an Early Intermediate on the Fold-Switching Pathway, Biochemistry 63 (2024) 2030-2039. https://doi.org/10.1021/acs.biochem.4c00258.
23. Wankowicz SA, de Oliveira SH, Hogan DW, van den Bedem H, Fraser JS, Ligand binding remodels protein side-chain conformational heterogeneity, Elife 11 (2022) https://doi.org/10.7554/eLife.74114.
24. Gonzalez-Higueras J, Freiberger MI, Galaz-Davison P, Parra RG, Ramirez-Sarmiento CA, A contact-based analysis of local energetic frustration dynamics identifies key residues enabling RfaH fold-switch, Protein Sci 33 (2024) e5182. https://doi.org/10.1002/pro.5182.